\documentclass[prc,twocolumn,showpacs,floatfix,nofootinbib]{revtex4}

\usepackage{graphicx}  %
\usepackage{bm}  %
\usepackage{amsmath,amssymb}

\newcommand{\beqn}{\begin{equation}}
\newcommand{\eeqn}{\end{equation}}
\newcommand{\bea}{\begin{eqnarray}}
\newcommand{\eea}{\end{eqnarray}}
\newcommand{\ba}{\begin{align}}
\newcommand{\ea}{\end{align}}

\newcommand{\vlowk}{V_{{\rm low}\,k}}

\newcommand{\fmi}{\, \text{fm}^{-1}}

\newcommand{\Hzero}{T_{\rm rel}}
\newcommand{\flow}{s}

\newcommand{\Vtwo}[2]{V_{#1#2}}
\newcommand{\Vthree}{V_{123}}
\newcommand{\Ttwo}[2]{T_{#1#2}}
\newcommand{\Trel}{T_{\rm rel}}

\begin{document}


\title{Similarity Renormalization Group\\ for Nucleon-Nucleon
            Interactions}

\author{S.K.\ Bogner}\email{bogner@mps.ohio-state.edu}
\author{R.J.\ Furnstahl}\email{furnstahl.1@osu.edu}
\author{R.J.\ Perry}\email{perry.6@osu.edu}

\affiliation{Department of Physics,
         The Ohio State University, Columbus, OH 43210}

\date{\today}
%

\begin{abstract}
The similarity renormalization group (SRG) is based on unitary
transformations that suppress off-diagonal matrix elements, forcing the
hamiltonian towards a band-diagonal form.
A simple SRG transformation 
applied to nucleon-nucleon interactions leads to 
greatly improved convergence properties while preserving observables,
and provides a method to consistently evolve many-body potentials
and other operators.
\end{abstract}

\smallskip
\pacs{21.30.-x,05.10.Cc,13.75.Cs}

\maketitle


Progress on the nuclear many-body problem has been hindered for decades
because
nucleon-nucleon (NN) potentials that reproduce elastic scattering
phase shifts typically exhibit strong short-range
repulsion as well as a strong tensor force.
This leads to strongly correlated many-body wave functions and
highly nonperturbative few- and many-body systems.
But recent work shows how a cutoff on relative momentum
can be imposed and evolved to  lower values
using renormalization group (RG) methods, thus eliminating the
troublesome high-momentum modes~\cite{Epelbaum,Vlowk2}. 
The evolved NN potentials
are energy-independent and preserve two-nucleon observables 
for relative momenta up to the cutoff.
Such potentials,
known generically as $\vlowk$,
are more perturbative and generate much less correlated
wave functions
\cite{Vlowk2,VlowkRG,Vlowk3N,Bogner_nucmatt,Bogner:2006tw,Bogner:2006vp},
vastly simplifying the many-body problem.
However, a full RG evolution of essential few-body potentials has not yet
been achieved.

An alternative path to decoupling high-momentum from
low-momentum physics is the similarity renormalization group (SRG),
which is based on unitary
transformations that suppress off-diagonal matrix elements, driving the
hamiltonian towards a band-diagonal form~\cite{Glazek:1993rc,Wegner:1994,%
Szpigel:2000xj,Roth:2005pd}. 
The SRG potentials are automatically energy independent and 
have the feature that high-energy phase shifts (and other high-energy
NN observables), while typically highly model dependent, are preserved,
unlike the case with $\vlowk$ as usually implemented.
Most important, 
\emph{the same
transformations renormalize all operators}, 
including many-body operators,
and the class of transformations can be tailored for effectiveness
in particular problems.

Here we make the first exploration of SRG for nucleon-nucleon
interactions, using a particularly simple choice of SRG transformation, 
which nevertheless works
exceedingly well.
We find the same benefits of $\vlowk$: more perturbative interactions
and lessened correlations, with improved convergence in few- and many-body
calculations.
The success of the SRG combined with advances in chiral effective field
theory (EFT) \cite{N3LO,N3LOEGM}
opens the door to the consistent  
construction and RG evolution of
many-body potentials and other operators. 

The similarity RG approach was developed independently by 
Glazek and Wilson \cite{Glazek:1993rc} 
and by Wegner \cite{Wegner:1994}.
We follow Wegner's formulation in terms of a flow equation
for the hamiltonian.
The initial 
hamiltonian in the center of mass $H = \Hzero + V$, where $\Hzero$ is the
relative kinetic energy, is transformed by the 
unitary operator $U(\flow)$ according to
\beqn
   H_\flow = U(\flow) H U^\dagger(\flow) \equiv \Hzero + V_\flow \ ,
\eeqn
where $\flow$ is the flow parameter.
This also defines the evolved
potential $V_\flow$, with $\Hzero$ taken to be independent of $\flow$.
Then $H_\flow$ evolves according to
\beqn
  \frac{dH_\flow}{d\flow}
    = [\eta(\flow),H_\flow] \ ,
\eeqn
with
\beqn
   \eta(\flow) = \frac{dU(\flow)}{d\flow} U^\dagger(\flow) 
          = -\eta^\dagger(\flow)
   \ .
\eeqn
Choosing $\eta(\flow)$ specifies the transformation.
Here we make perhaps the simplest choice \cite{Szpigel:2000xj},
\beqn
  \eta(\flow) =  [\Hzero, H_\flow]
  \label{eq:choice}
   \ ,
\eeqn
which gives the flow equation,
\beqn
  \frac{dH_\flow}{d\flow} 
  =  [ [\Hzero, H_\flow], H_\flow] \ .
  \label{eq:commutator}
\eeqn 
Other choices will be studied elsewhere \cite{Bogner:2007a}. 
 
\begin{figure*}[htb]
  \includegraphics*[width=3.2in]{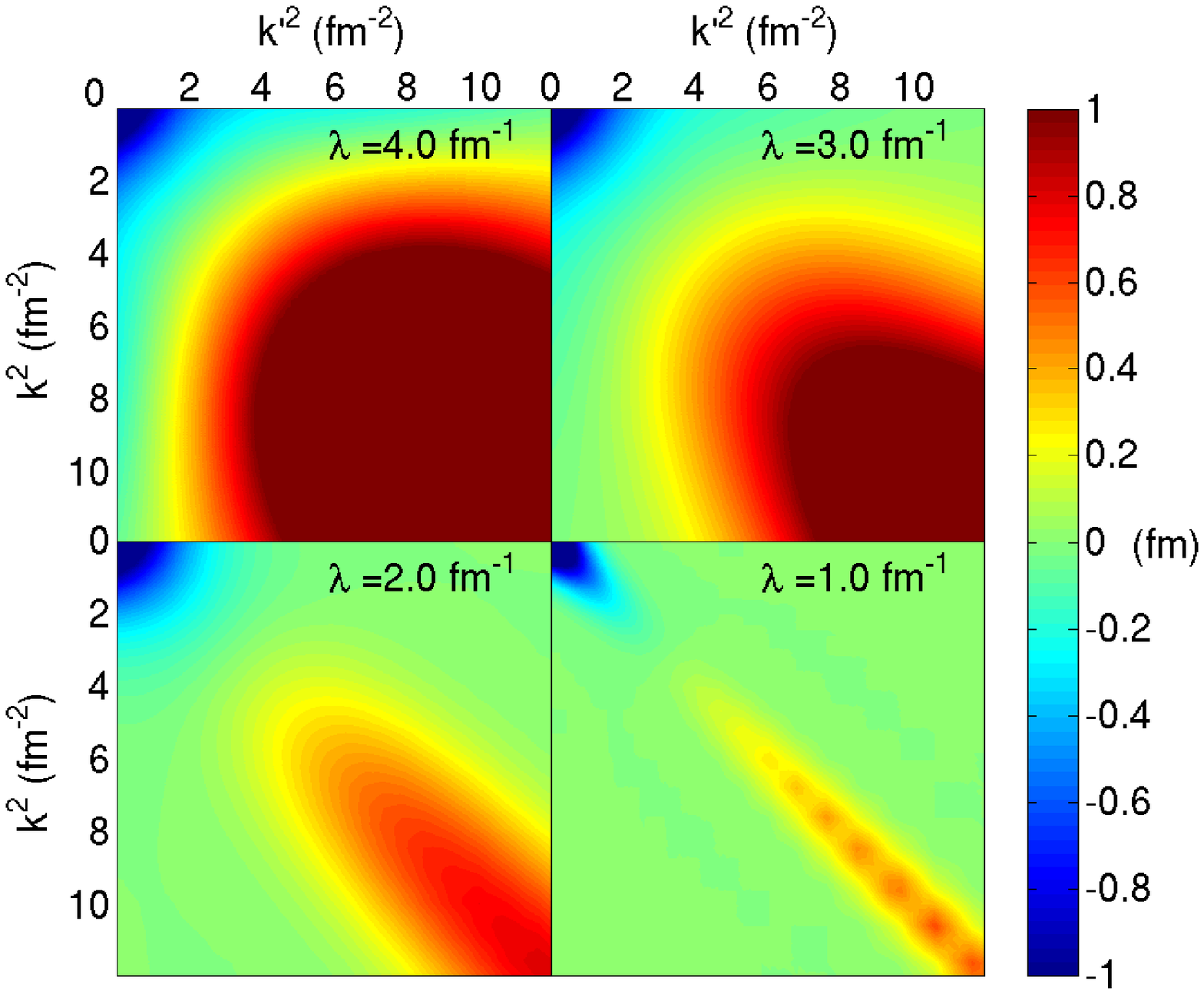}
  \hspace*{.3in}
  \includegraphics*[width=3.2in]{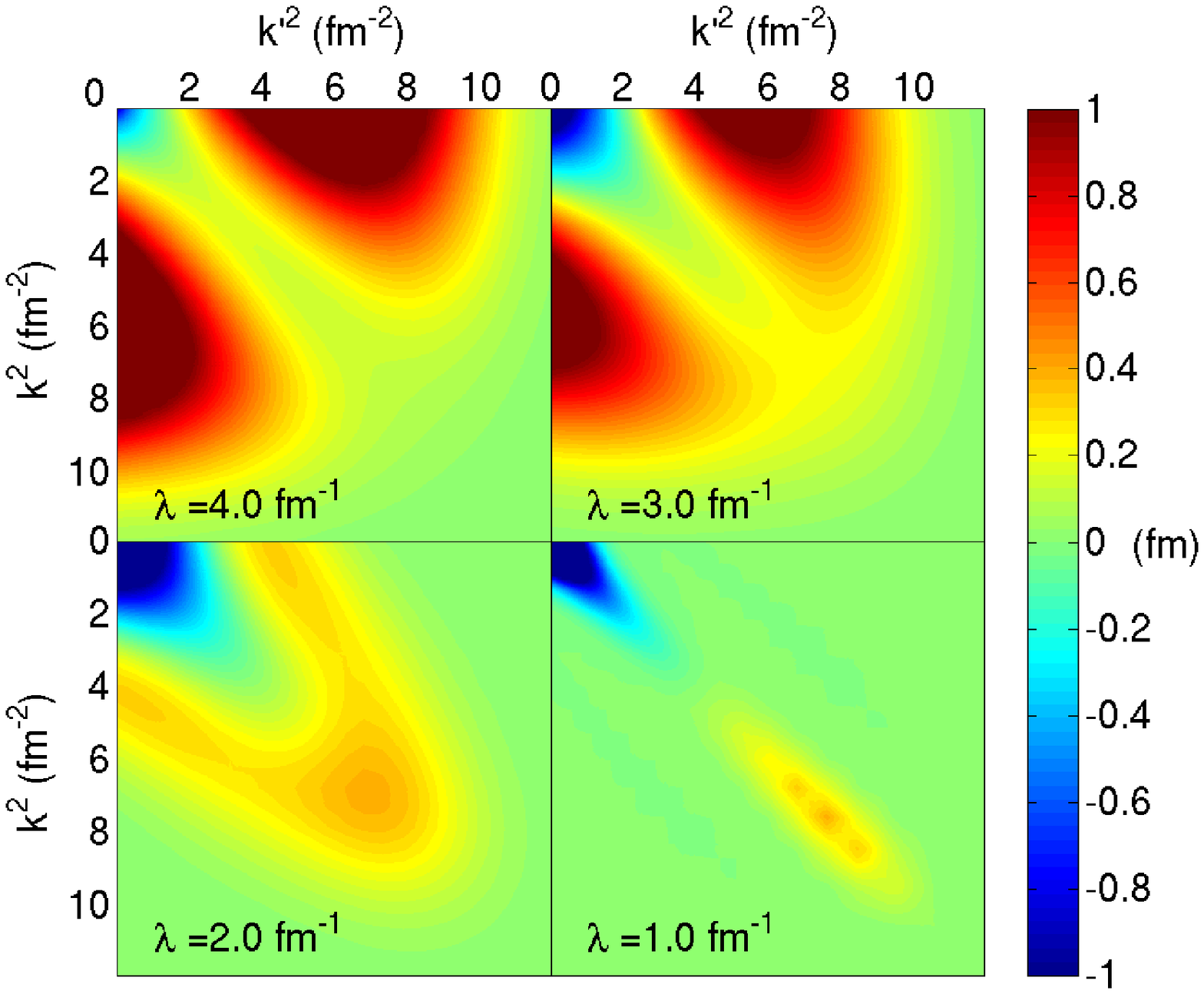}
  \vspace*{-.1in}
  \caption{(Color online) Contour plots of $V_\flow(k,k')$ illustrating
  the evolution with $\lambda \equiv s^{-1/4}$
  for $^1$S$_0$ (left) and $^3$S$_1$ (right).  
  The initial potential on the left is a chiral N$^3$LO 
  potential
  with a 600\,MeV cutoff \cite{N3LO} and on the right is an N$^3$LO
  potential with a 550\,MeV cutoff on
  the Lippmann-Schwinger equation and a 600\,MeV cutoff on a regularized
  spectral representation of two-pion exchange \cite{N3LOEGM}.}
  \label{fig:vsrg}
\end{figure*}
  
\begin{figure*}[htb]
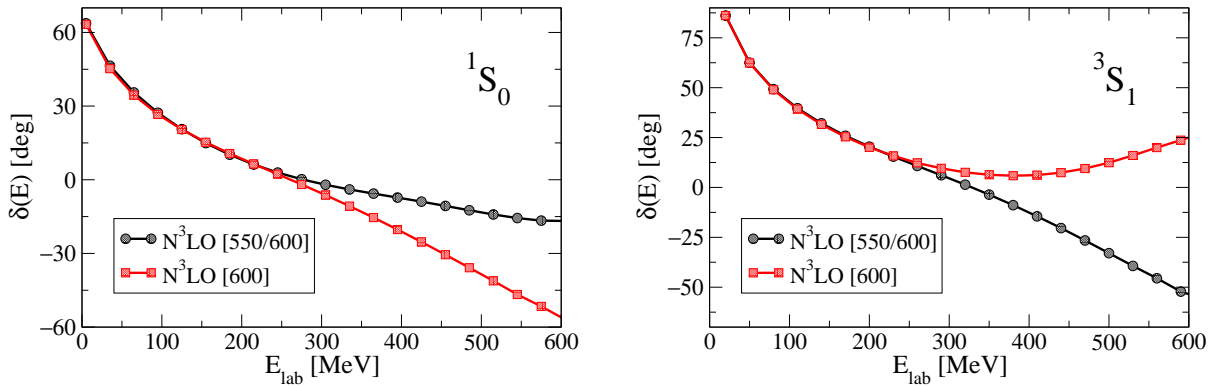

  \includegraphics*[width=3.0in]{figure2a}
  \hspace*{.2in}
  \includegraphics*[width=3.0in]{figure2b}
  \vspace*{-.1in}
  \caption{(Color online) 
  S-wave phase shifts from the two chiral EFT N$^3$LO potentials
  from Fig.~\ref{fig:vsrg}.
  For each initial potential,
  the phase shifts for different $\lambda$ agree to within
  the widths of the lines at all energies shown.}
  \label{fig:phases}
\end{figure*}
 
For any given partial wave
in the space of relative momentum NN states, 
Eq.~(\ref{eq:commutator}) means that 
the potential in momentum space evolves
as (with normalization
so that $1 = \frac{2}{\pi}\int_0^\infty|q\rangle q^2\,dq \langle q |$
and in units where $\hbar^2/M = 1$),
\bea
  \frac{dV_\flow(k,k')}{d\flow} &=&
    - (k^2 - k'{}^2)^2 \, V_\flow(k,k')
    \nonumber 
    \\* & &  \null
    + \frac{2}{\pi}\int_0^\infty\! q^2\,dq\
      (k^2 + k'{}^2 - 2q^2)\,
    \nonumber   
    \\* & & \qquad\qquad\null \times
      V_\flow(k,q)\, V_\flow(q,k')
      \ .
      \label{eq:diffeq}
\eea
(The additional matrix structure of $V_\flow$ in coupled channels
such as $^3$S$_1$--$^3$D$_1$ is implicit.)
For matrix elements far from the diagonal, the first term
on the right side of Eq.~(\ref{eq:diffeq}) evidently dominates and 
exponentially suppresses
these elements as $\flow$ increases.
The parameter $\lambda \equiv s^{-1/4}$ provides a measure of the
spread of off-diagonal strength.
While further analytic analysis is possible, we instead turn to a numerical
demonstration that the flow toward the diagonal is a general result.  
By discretizing the relative momentum space on a grid of gaussian integration
points, 
we obtain a simple (but nonlinear) system of first-order
coupled differential equations,
with the boundary condition that $V_\flow(k,k')$ at the initial
$\flow$ (or $\lambda$) is equal to the initial potential.

The evolution of the hamiltonian 
according to Eq.~(\ref{eq:diffeq}) as $\flow$ increases 
(or $\lambda$ decreases)
is illustrated in Fig.~\ref{fig:vsrg},
using two initial chiral EFT potentials \cite{N3LO,N3LOEGM}.  
On the left in Fig.~\ref{fig:vsrg}
is $^1$S$_0$ starting from the harder (600\,MeV cutoff)
potential from Ref.~\cite{N3LO}, which has
significant strength near the high-momentum diagonal, and on the right is
the S-wave part of the $^3$S$_1$--$^3$D$_1$ coupled channel starting
from one of the potentials from Ref.~\cite{N3LOEGM},
which has more far off-diagonal strength initially and comparatively weaker
higher-momentum strength on the diagonal.  
The initial momentum-space potential differs significantly among
interactions that are phase-equivalent up to the NN inelastic threshold,
but these examples show characteristic features of the evolution in
$\lambda$. 
In particular, we see a systematic suppression of off-diagonal strength,
as anticipated, with the width of the diagonal scaling as $\lambda^2$.
The same behavior is observed when evolving from conventional high-precision
NN potentials, such as Argonne $v_{18}$, or other (softer) chiral
potentials \cite{Bogner:2007a,Bogner:2007jb}.

Since the SRG transformation is unitary, observables
are unchanged at \emph{all} energies, up to numerical errors.
This is shown by 
Fig.~\ref{fig:phases}, in which phase shifts for the two chiral EFT 
potentials are plotted, including the values at high energies where they are
not constrained by data (above $E_{\rm lab} = 300\,$MeV).
For a given potential, there is no visible variation with $\lambda$.
Similarly, the binding energy and asymptotic normalizations for the deuteron
are independent of $\lambda$ \cite{Bogner:2007jb}.

\begin{figure}[ht]
\includegraphics*[width=3.2in]{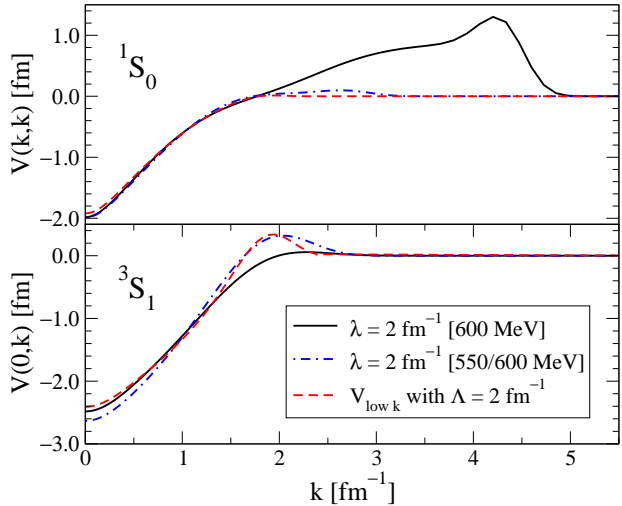}
\vspace*{-.1in}
\caption{(Color online) Matrix elements of the evolved SRG potentials at 
$\lambda = 2\,\fmi$
for $^1$S$_0$ (top, diagonal elements) and  $^3$S$_1$ 
(bottom, off-diagonal elements)
for the same initial potentials as in Fig.~\ref{fig:vsrg}.
Also shown is the $\vlowk$ potential with a smooth (exponential)
regulator for momentum cutoff $\Lambda = 2\,\fmi$, evolved from 
the two potentials (600 MeV above and
550/600 MeV below).}
\label{fig:compare}
\end{figure}

As $\lambda$ is lowered, different initial potentials flow to 
similar forms at low momentum while remaining distinct
at higher momentum.
The low-momentum parts also become similar to $\vlowk$ potentials.
These observations are illustrated  in Fig.~\ref{fig:compare}
for two particular slices of the
potentials from Fig.~\ref{fig:vsrg}. 
They will be explored in much greater detail in Ref.~\cite{Bogner:2007a}.
 
We can quantify the perturbativeness of the potential as we evolve
to lower $\lambda$ by using the eigenvalue analysis introduced long ago
by Weinberg \cite{Weinberg} and recently applied in an analysis of
$\vlowk$ potentials \cite{Bogner:2006tw}. 
Consider the operator Born series for the $T$-matrix at energy $E$
(for simplicity we assume $E \leq 0$):
\beqn
 T(E) = V_\flow + V_\flow \frac{1}{E-\Hzero} V_\flow + \cdots
\eeqn 
By finding the eigenvalues and eigenvectors of 
\beqn
  \frac{1}{E - \Hzero} V_\flow | \Gamma_\nu \rangle
    = \eta_\nu(E) | \Gamma_\nu \rangle
    \ ,
\eeqn
and then acting with $T(E)$ on the eigenvectors,
\beqn
 T(E) | \Gamma_\nu \rangle
   =  V_\flow | \Gamma_\nu \rangle
   (1 + \eta_\nu + \eta_\nu^2 + \cdots)
   \ ,
\eeqn
it follows that nonperturbative behavior at energy $E$ is signaled
by one or more eigenvalues with $|\eta_\nu(E)| \geq 1$ \cite{Weinberg}.
(See Ref.~\cite{Bogner:2006tw} for a more detailed discussion in
the context of evolving $\vlowk$ potentials.)

\begin{figure}[ht]
\includegraphics*[width=2.6in]{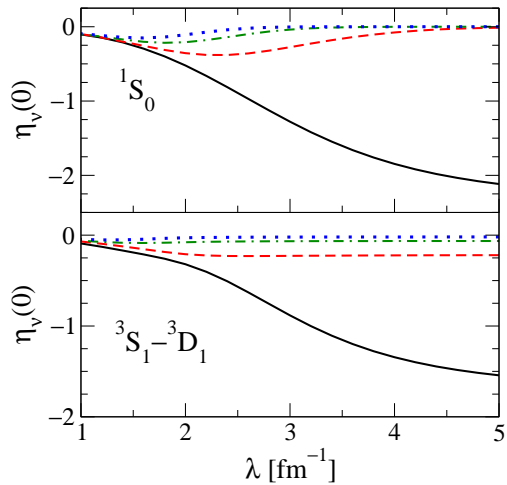}
\vspace*{-.1in}
\caption{(Color online) The largest
repulsive Weinberg eigenvalues as a function of $\lambda$
in the $^1$S$_0$ channel and the $^3$S$_1$--$^3$D$_1$ coupled channel
for the same initial potentials as in Fig.~\ref{fig:vsrg}.}
\label{fig:weinberg}
\end{figure}

It suffices for our purposes to consider a single energy (e.g., $E=0$),
and to consider only the negative eigenvalues, which are associated with the
short-range repulsion.
Weinberg eigenvalues at zero energy
are shown as a function of
$\lambda$ in  Fig.~\ref{fig:weinberg}
for the $^1$S$_0$ channel and the $^3$S$_1$--$^3$D$_1$ coupled
channel.
In both channels, the large negative eigenvalues at large
$\lambda$ reflect the repulsive core of the initial potentials.
They rapidly evolve to small values as $\lambda$ 
decreases to $2\,\mbox{fm}^{-1}$ and below, as also observed with the 
corresponding $\vlowk$
evolution \cite{Bogner:2006tw}.  However, the intermediate
increase for the sub-leading eigenvalues in $^1$S$_0$ is a new feature 
of the SRG that merits further study.

\begin{figure}[bht]
\includegraphics*[width=3.0in]{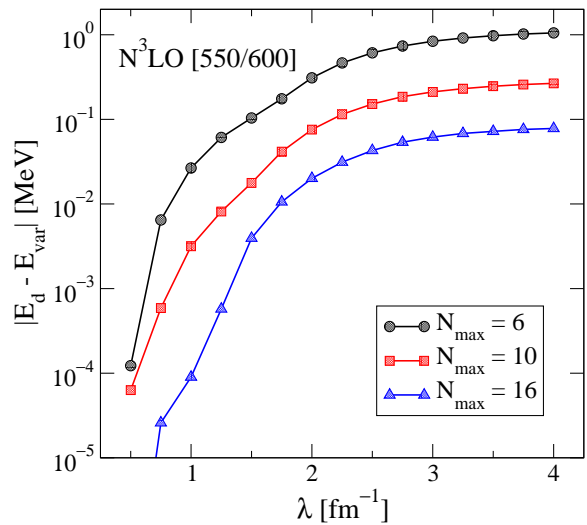}
\vspace*{-.1in}
\caption{(Color online) The absolute error vs.\ $\lambda$
of the predicted deuteron binding energy 
from a variational calculation in a fixed-size basis of harmonic
oscillators ($N_{\rm max} \hbar\omega$ excitations).  
The initial potential is from Ref.~\cite{N3LOEGM}.}
\label{fig:deuteron_convergence}
\end{figure}

\begin{figure}[!tbh]
\includegraphics*[width=3.1in]{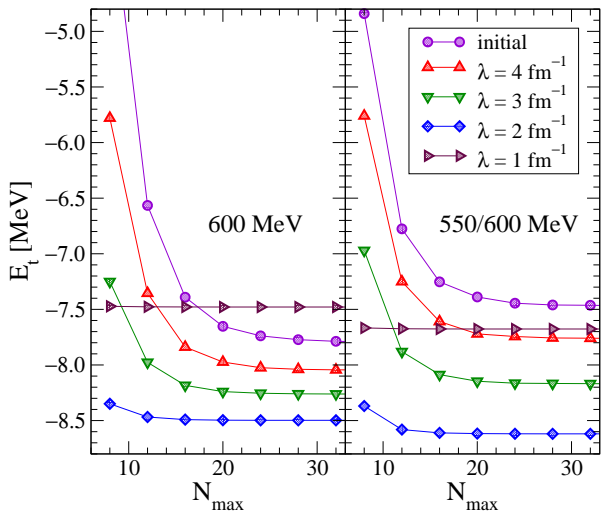}
\vspace*{-.1in}
\caption{(Color online) The variational binding energy for selected $\lambda$
of the triton with two-nucleon interactions only, 
as a function of the size of the harmonic oscillator space
($N_{\rm max} \hbar\omega$ excitations),
for the same initial potentials as in Fig.~\ref{fig:vsrg}.}
\label{fig:triton_convergence}
\end{figure}

The more perturbative potentials at lower $\lambda$ 
induce weaker short-range correlations in
few- and many-body wave functions, which in turn leads to
greatly improved convergence in variational calculations.
This is illustrated via calculations of the binding energy
of the deuteron and
triton by diagonalization in a harmonic oscillator basis,
as shown in Figs.~\ref{fig:deuteron_convergence} and
\ref{fig:triton_convergence}.
For a fixed basis size, a more accurate estimate is obtained
with smaller $\lambda$ or, conversely, at fixed $\lambda$ the
convergence with basis size becomes more rapid.
The improvement in convergence is similar to that found
with smoothly regulated $\vlowk$ potentials evolved
from chiral N$^3$LO potentials \cite{Bogner:2006vp}.
At finite density, analogous effects led to perturbative 
behavior in nuclear matter for $\vlowk$ potentials \cite{Bogner_nucmatt}.
Results for the G matrix
support a similar conclusion for the SRG 
potentials (see the website in Ref.~\cite{Bogner:2007a} for 
pictures).

In Fig.~\ref{fig:triton_convergence}, the
calculations for different $\lambda$ converge to different values for
the binding energy of the triton. This reflects the 
contributions of the omitted (and evolving) three-body interaction. 
The evolution with $\lambda$ of the binding energy with NN
interactions only, which is also the evolution of the net three-body
contribution, follows a similar pattern to that seen with 
$\vlowk$ \cite{Vlowk3N,Bogner:2006vp}:
a slow decrease as $\lambda$ decreases, reaching a minimum for $\lambda$
between
$1.5\,\fmi$ and $2\,\fmi$, and then a rapid increase.

The consistent RG evolution of few-body interactions is an important unsolved
problem for low-momentum potentials. In $\vlowk$ calculations to date, an
approximate evolution is made by fitting the leading chiral EFT
three-body force at each cutoff while evolving the two-body interaction
exactly \cite{Vlowk3N}. 
Generalizing the RG evolution to include the three-body interaction
is, at least, technically challenging.
This is because 
the machinery used to construct
$\vlowk$ requires the solution of the full
three-nucleon  problem (i.e., bound state wave functions plus all
scattering wave functions in all breakup channels) to consistently
evolve the three-nucleon interactions.  

In contrast, the SRG evolution follows by directly
applying Eq.~(\ref{eq:commutator}) in the three-particle space.
To show the basic idea, we adopt a notation in which 
$\Vtwo12$ means the two-body interaction between particles 1
and 2 while $\Vthree$ is the irreducible three-body potential.
In standard treatments of the nuclear three-body problem
(for example, Ref.~\cite{fewbody}), the relative kinetic energy for equal
mass particles with mass $m$ is decomposed
as (with the total momentum ${\bf K} = \sum_i {\bf k}_i = 0$):
\beqn
  \Trel = \sum_i \frac{{\bf k}_i^2}{2m}
    = \frac{{\bf p}_l^2}{m} + \frac{3{\bf q}_l^2}{4m}
    \ , \qquad
    l = 1,2,3
    \;,
\eeqn 
where $l$ denotes which set of Jacobi coordinates are being used
to describe the relative motion
(i.e., $l=1$ means 
${\bf p}_1 = \frac{1}{2}({\bf k}_2 - {\bf k}_3)$
and 
${\bf q}_1 = \frac{2}{3}[{\bf k}_1 
  - \frac{1}{2}({\bf k}_2 + {\bf k}_3)]$,
and so on).
In the notation here, 
\beqn
   \frac{{\bf p}_1^2}{m} \leftrightarrow \Ttwo23
   \qquad \mbox{and} \qquad
   \frac{3{\bf q}_1^2}{4m} \leftrightarrow T_1 \;,
\eeqn
and similarly for $l=2$ and $l=3$.

Now we
start with the hamiltonian including up to three-body interactions
(keeping in mind that higher-body interactions will be induced
as we evolve in $s$ but will not contribute to three-body systems):
\beqn
  H_s = \Trel + \Vtwo12 + \Vtwo13 + \Vtwo23 + \Vthree
    \equiv \Trel + V_s
    \;.
  \label{eq:H}
\eeqn 
(Note: all of the potentials depend implicitly on $s$.)
The relative kinetic energy operator $\Trel$ can be decomposed 
in three ways:
\bea
  \Trel = \Ttwo12 + T_3 = \Ttwo13 + T_2 = \Ttwo23 + T_1
  \;,
  \label{eq:Trel}	
\eea 
and $T_i$ commutes with $\Vtwo{j}{k}$,  
\beqn
   [T_3, \Vtwo12] = [T_2, \Vtwo13] = [T_1, \Vtwo23] = 0
   \ ,
   \label{eq:Tcomm}
\eeqn
so the commutators of $\Trel$ with the two-body potentials become
\beqn
  [\Trel,\Vtwo12] = [\Ttwo12,\Vtwo12] \;, 
  \label{eq:comms}
\eeqn
and similarly for $\Vtwo13$ and $\Vtwo23$. 

Since we define $\Trel$ to be independent of $s$, 
the SRG equation (\ref{eq:commutator}) 
for the three-body Hamiltonian $H_s$ simplifies to
\beqn
  \frac{dV_s}{ds} = 
 \frac{d\Vtwo12}{ds} + \frac{d\Vtwo13}{ds} + \frac{d\Vtwo23}{ds}
  + \frac{d\Vthree}{ds}
  = 
  [[\Trel, V_s], H_s] \;,
  \label{eq:dVs}
\eeqn 
with $V_s$ from Eq.~(\ref{eq:H}).
The corresponding equations for each of the two-body potentials
(which are completely determined by their evolved matrix elements in the
two-body systems, e.g., on a gaussian momentum grid) are
\bea
   \frac{d\Vtwo12}{ds} &=& [[\Ttwo12, \Vtwo12], (\Ttwo12 + \Vtwo12)] 
   \ , \label{eq:Vtwoa} 
\eea
and similarly for $\Vtwo13$ and $\Vtwo23$.
After
expanding Eq.~(\ref{eq:dVs}) using Eq.~(\ref{eq:H}) and the different
decompositions of $\Trel$,
it is straightforward to show using the equations for the two-body potentials 
that the
derivatives of two-body potentials
on the left side cancel precisely with terms on the right
side, leaving
\bea
 \frac{d\Vthree}{ds} &=&
 [[\Ttwo12,\Vtwo12], (T_3 + \Vtwo13 + \Vtwo23 + \Vthree)]
  \nonumber \\ & & \null +
 [[\Ttwo13,\Vtwo13], (T_2 + \Vtwo12 + \Vtwo23 + \Vthree)]
  \nonumber \\ & & \null +
 [[\Ttwo23,\Vtwo23], (T_1 + \Vtwo12 + \Vtwo13 + \Vthree)]
  \nonumber \\ & & \null +
  [[\Trel,\Vthree],H_s]
  \ .
  \label{eq:diffeqp}
\eea
The importance of these cancellations is that they eliminate the
``dangerous'' delta functions, which make setting up the integral
equations for the three-body system problematic \cite{fewbody}. 
We emphasize that the $s$-dependence of the two-body interactions
on the right side of Eq.~\ref{eq:diffeqp} is completely specified by
solving the two-body problem in Eq.~(\ref{eq:diffeq}). This is in
contrast to RG methods that run a cutoff on the total energy of the
basis states (e.g., Lee-Suzuki and Bloch-Horowitz techniques). Such
methods generate ``multi-valued'' two-body interactions, in the sense
that the RG evolution of two-body operators in $A>2$ systems depends 
non-locally on the excitation energies of the unlinked spectator 
particles \cite{multivalued_ls,multivalued_bh}. 
 
Further simplications of Eq.~(\ref{eq:diffeqp}) follow from symmetrization
and applying the Jacobi identity, but this form is sufficient to
make clear that there are no disconnected pieces.
The problem is thus reduced to the technical implementation of
a momentum-space decomposition analogous to Eq.~(\ref{eq:diffeq}). 
Note in particular
that the evolution of $\Vthree$ is carried out without ever
having to solve a bound-state or scattering problem.
We have verified that this formalism does generate three-body 
interactions that leave eigenvalues invariant for
simple model hamiltonians,
such as a two-level system of bosons.
Work is in progress on proof-of-principle tests using one-dimensional
many-body systems (to avoid angular momentum complications) and
the three-dimensional nuclear problem will be tackled soon.

In summary, 
the SRG applied to nucleon-nucleon potentials
works as advertised even for a simple
choice of transformation, driving the hamiltonian (in momentum
space) towards the diagonal, making it more perturbative and more
convergent in few-body calculations.
There is much to explore, such as
the nature of the decoupling of high- and low-energy physics implied by
Fig.~\ref{fig:vsrg} (see Ref.~\cite{Bogner:2007jb} for demonstrations
of decoupling and the evolution of an operator in the
two-body space) and whether
other choices of $\eta(s)$ instead of 
Eq.~(\ref{eq:choice}) could be more effective
in making the hamiltonian diagonal. 
For example, the replacement $\Hzero \rightarrow H_d$, where $H_d$ is the
diagonal part of the hamiltonian,
or some function of $\Hzero$
are easily implemented.
Most important is the consistent evolution
of nuclear three-body operators~\cite{Bogner:2007a}.


\vspace*{.1in}

\begin{acknowledgments}
We thank A. Schwenk for useful comments. 
This work was supported in part by the National Science 
Foundation under Grant No.~PHY--0354916 and  
the Department of Energy under Grant No. DE-FC02-07ER41457.
\end{acknowledgments}


\begin{thebibliography}{99}

\bibitem{Epelbaum} E. Epelbaum, W. Gl\"ockle, A. Kr\"uger 
and U.G. Mei{\ss}ner, Nucl. Phys. \textbf{A645}, 413 (1999).


\bibitem{Vlowk2} S. K.\ Bogner, T. T. S.\ Kuo and A.\ Schwenk, 
Phys.\ Rept.\ \textbf{386}, 1 (2003).

\bibitem{VlowkRG} S. K.\ Bogner, A.\ Schwenk, T. T. S.\ Kuo and G. E.\ Brown,
nucl-th/0111042.

\bibitem{Vlowk3N} A.\ Nogga, S. K.\ Bogner and A.\ Schwenk, Phys.\ Rev.\
C \textbf{70},  061002(R) (2004).

\bibitem{Bogner_nucmatt} S. K.\ Bogner, A.\ Schwenk, R. J.\ Furnstahl and 
A.\ Nogga, Nucl.\ Phys.\ \textbf{A763}, 59 (2005).


\bibitem{Bogner:2006tw} S. K.\ Bogner, R. J.\ Furnstahl, S.\ Ramanan and 
A.\ Schwenk, Nucl.\ Phys.\ \textbf{A773}, 203 (2006).

\bibitem{Bogner:2006vp}
  S. K.~Bogner, R. J.~Furnstahl, S.~Ramanan and A.~Schwenk,
  Nucl.\ Phys.\ \textbf{A784}, 79 (2007).

\bibitem{Glazek:1993rc}
  S. D.~Glazek and K. G.~Wilson,
  Phys.\ Rev.\ D {\bf 48}, 5863 (1993);
%
  Phys.\ Rev.\ D {\bf 49}, 4214 (1994).

\bibitem{Wegner:1994}
  F. Wegner, Ann.\ Phys.\ (Leipzig) {\bf 3}, 77 (1994).

\bibitem{Szpigel:2000xj}
  S.~Szpigel and R.~J.~Perry,
  in \textit{Quantum Field Theory, A 20th Century Profile}
  ed.\ A.N.\ Mitra, (Hindustan Publishing Com., New Delhi, 2000),  
  arXiv:hep-ph/0009071.

\bibitem{Roth:2005pd}
  An alternative non-RG use of unitary transformations to reduce
  correlations in many-body wave functions is described in
  R.~Roth, H.~Hergert, P.~Papakonstantinou, T.~Neff and H.~Feldmeier,
  Phys.\ Rev.\ C {\bf 72}, 034002 (2005), and references therein.
  The relationship of this approach to the SRG is described in
  H.~Hergert and R.~Roth, Phys.\ Rev.\ C {\bf 75}, 051001(R) (2007).
  

\bibitem{N3LO} D. R.\ Entem and R.\ Machleidt, Phys.\ Rev.\ C \textbf{68}, 
   041001(R) (2003).

\bibitem{N3LOEGM} E.\ Epelbaum, W.\ Gl\"ockle and U. G.\ Mei{\ss}ner,
Nucl.\ Phys.\ \textbf{A747}, 362 (2005).
  
\bibitem{Bogner:2007a}
  S. K. Bogner, R. J. Furnstahl, and R. J. Perry, in preparation.
  See also http://www.physics.ohio-state.edu/$\sim$ntg/srg/ for
  documentary examples of the SRG applied to NN potentials.

\bibitem{Bogner:2007jb}
  S.~K.~Bogner, R.~J.~Furnstahl, R.~J.~Perry and A.~Schwenk,
  Phys.\ Lett.\ B {\bf 649}, 488 (2007).
    
\bibitem{Weinberg} S.\ Weinberg, Phys.\ Rev.\ \textbf{131}, 440 (1963).


\bibitem{fewbody}W. Gl\"ockle, \textit{The Quantum Mechanical Few-Body Problem}
   (Springer-Verlag, Berlin, 1983). 
\bibitem{multivalued_ls}P. Navratil, B.~R.~Barrett, Phys.\ Rev.\ C \textbf{54},
 2986 (1996).
\bibitem{multivalued_bh}T.~C.~Luu, S.~K.~Bogner, W.~C.~Haxton, and P.~Navratil, 
Phys.\ Rev.\ C \textbf{70}, 014316 (2004).
 

\end{thebibliography}
\end{document}